When 2D is not enough, go for an extra dimension

Thierry Rabilloud 1,2,3

1: CNRS, Laboratory of Chemistry and Biology of Metals (LCBM), UMR 5249, Grenoble, France
2: Univ. Grenoble Alpes, LCBM, Grenoble, France
3: CEA, iRTSV/LCBM, Grenoble France

Postal address
Pro-MD team, UMR CNRS-CEA-UJF 5249, iRTSV/LCBM, CEA Grenoble, 17 rue des martyrs, 38054 Grenoble CEDEX 9, France

email: thierry.rabilloud@cea.fr

Abstract
The use of an extra SDS separation in a different buffer system provide a technique for deconvoluting 2D gel spots made of several proteins [Colignon et al. Proteomics, 2013, 13: xxx-yyy]. This technique keeps the quantitative analysis of the protein amounts and combines it with a strongly improved identification process by mass spectrometry, removing identification ambiguities in most cases. In some favorable cases, post-translational variants can be separated by this procedure. This versatile and easy to use technique is anticipated to be a very valuable addition to the toolbox used in 2D gel-based proteomics.

Main text

Because of its robustness, capacity to handle large sample series, easy interface with many other biochemical techniques and above all its unique ability to analyse complete proteins, 2D gel electrophoresis is still a relevant approach in many proteomic studies [1, 2]. In most cases, 2D gel electrophoresis is used as a first quantitative screening process to select the spots which abundance change upon the biological phenomenon of interest. It is then necessary to identify the proteins present in these modulated spots to obtain a quantitative biochemical view of the molecular events at play in the biological phenomenon of interest.
For many years, i.e. from 1990 to 2005, this has been a straightforward process, as the protein analysis techniques, namely Edman's sequencing and then mass spectrometry, always rendered one protein per electrophoretic 2D spot. However, with the ever increasing sensitivity of mass spectrometers, this equation is less and less true with the proportion of singulets (one protein per spot) going down, from 70% in 2005 [3] to 50% in 2013 [4]. This does not disqualify 2D gel-based proteomics per se, as previously stated [5], as long as it is possible to correlate with good confidence the variation in a spot volume with the variation of one protein. Using a reduction ad absurdum, all the historical data using low sensitivity methods able to identify a moderately complex mixture of proteins, whether Edman's sequencing [6, 7], or tandem mass spectrometry [8, 9] point to the fact that almost all 2D spots are made of a major protein, which explains the quantitative variations in staining, and of minor components that do not play any role in the staining variation but are real components of the protein spot.

This fact should not come as a surprise, when integrating on the one hand the resolving power of 2D gels and on the other hand the dynamic range of the proteome and the number of different protein species present in any complex biological sample.

Thus, the name of the game is to identify this major component in the protein spots. As shown in figure 1 and table 1, this is sometimes quite straightforward (spots 1 to 3) and sometimes almost impossible (spots 4 to 8) from the simple MS/MS output. Several approaches can be designed to circumvent this major problem. The simplest one is to analyse less and less protein in the spots, so that the mass spectrometer will find only the most abundant component. This can be achieved by analyzing small silver-stained spots, with the major risk that many spots will no longer show any identification.

A much more powerful and elegant approach is to combine the peptide by peptide quantitative analysis of SILAC with the ability of 2D gels to resolve complete proteins, as exemplified by a recent work on HeLa cells [4]. However, not all biological systems are easily amenable to SILAC labelling, which is in addition a costly procedure.

All in all, the ideal solution would be to be able to deconvolute the 2D spots into their individual proteic components in a quantitative way. This would allow to check which component(s) account for the quantitative variation in staining while making the mass spectrometric identification unambiguous again. This is exactly what is achieved by the third dimension electrophoresis described by Colignon et al. [10], in which the 2D spots are excised and re-electrophoresed on a different gel system to resolve them into individual components.

Three-dimensional electrophoresis has been described in the past, but in most cases the third dimension is carried before the conventional IEF-SDS separation [11, 12] and not after it as in the Colignon paper. Consequently, these approaches need to know upfront how to separate the proteins, which is seldom the case in most proteomic studies. Perhaps the most impressive 3D electrophoresis is the gel cube [13], which uses IEF as one separation dimension and two different SDS systems in the other two. While theoretically equivalent to the Colignon setup, the gel cube is much more cumbersome to handle and clearly not as straightforward as an approach which can be carried out on minigels and on only the spots that need it. The only approach described in the literature that can be compared with the Colignon setup is the one described by Vanfleteren [14], but it used a very specialized electrophoretic system which may not be applicable to all proteins.

The few examples shown in the Colignon paper demonstrate both the power and the limitation of the method. As the third dimension is also a SDS electrophoresis, some proteins that comigrate in 2D gels will still comigrate in the third dimension, leading to multiple identifications in mass spectrometry. Despite this intrinsic limitation, the technique works surprisingly well in its ability to deconvolute 2D gels spots, and the well-known versatility of SDS electrophoresis calls for a very wide scope of application, with very few proteins intractable to the third dimension. In addition, it seems that the third dimension may be sometimes able to separate post-translational variants, which further adds to the attractivity of the technique.
In summary, this technique is easy to implement, cheap, and is likely to bring in most cases the extra separation that is more and more needed to interface safely 2D gels with the more and more sensitive mass spectrometers.

Footnote: the author has declared no conflict of interest

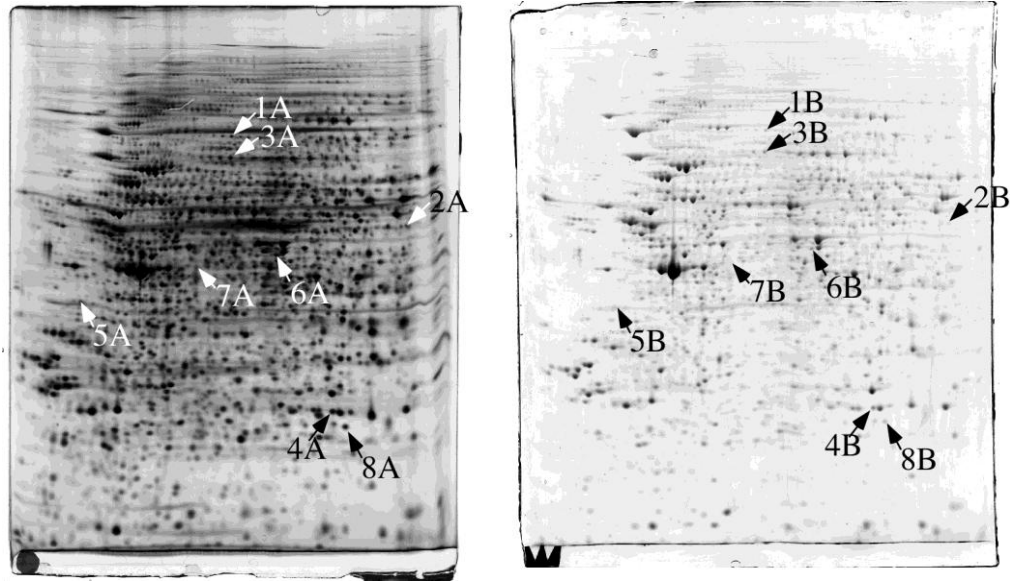

Figure 1: comparison of silver-stained and Coomassie blue-stained gels.
A whole cell extract of RAW264 murine macrophage cell line was separated by 2D gel electrophoresis (linear pH gradient ranging from 4 to 8 in the first dimension, 10% T gel in the second dimension).

Left panel: one hundred micrograms of proteins loaded on the first dimension gel, detection by silver staining

Right panel: five hundred micrograms of proteins loaded on the first dimension gel, detection by colloidal Coomassie Blue

Equivalent spots were excised on the two gels, digested with trypsin, and the resulting peptides analyzed by tandem mass spectrometry on an ion trap instrument. The resulting data are presented on table 1.

Table 1: identification data for the spots excised from the gels shown in Figure 1

Spots 1 to 3 illustrate easy cases where no ambiguity is encountered, whereas spots 4 to 8 illustrate difficult cases where no straightforward identification can be made when supra-optimal amounts of proteins are present in the spots.

| spot Nb. | protein name | Swissprot accession number | MW | Nb. Unique peptides | sequence coverage |
|---|---|---|---|---|---|
| 1A | Mitofilin | Q8CAQ8 | 83901,1 | 2 | 2,64% |
| 1B | Mitofilin | Q8CAQ8 | 83901,1 | 33 | 52,40% |
|  | ATP synthase subunit alpha | Q03265 | 59754,1 | 1 | 2,71% |
| 2A | ATP synthase subunit alpha | Q03265 | 59754,1 | 8 | 15,90% |
| 2B | ATP synthase subunit alpha | Q03265 | 59754,1 | 28 | 54,60% |
|  | Glutamate dehydrogenase 1 | P26443 | 61417,4 | 2 | 3,76% |
|  | Irg1 | P54987 | 53759,6 | 1 | 2,05% |
|  | Serbp1 | Q9CY58 | 44754,6 | 1 | 3,93% |
|  | Ripk3 | Q9QZL0 | 53336,4 | 1 | 2,88% |
|  | Ugp2 | Q91ZJ5 | 56925,9 | 1 | 2,36% |
| 3A | Moesin | P26041 | 67821,8 | 2 | 2,60% |
| 3B | Moesin | P26041 | 67768,8 | 27 | 40,20% |
|  | Glycine--tRNA ligase | Q9CZD3 | 81879,1 | 15 | 19,50% |
|  | Beta-glucuronidase | P12265 | 74195,7 | 3 | 4,63% |
|  | CTP synthase 1 | P70698 | 66690,7 | 3 | 4,74% |
|  | PDI A4 | P08003 | 71984,4 | 3 | 5,17% |
| 4A | Hprt1 | P00493 | 24571,3 | 3 | 16,50% |
| 4B | Hprt1 | P00493 | 24571,3 | 5 | 27,10% |
|  | Triosephosphate isomerase | P17751 | 32191,3 | 4 | 19,40% |
|  | Flavin reductase (NADPH) | Q923D2 | 22196,7 | 2 | 11,70% |
|  | GTP-binding nuclear protein | P62827 | 24427,3 | 2 | 11,10% |
|  | dtd1 | Q9DD18 | 23232,3 | 1 | 7,18% |
|  | GSH S-transferase Mu 5 | P48774 | 26636,5 | 1 | 4,02% |
|  | Pcmt1 | P23506 | 24641,9 | 1 | 4,85% |
| 5A | V-type H+ ATPase sub. d 1 | P51863 | 40302,8 | 2 | 6,84% |
| 5B | V-type H+ ATPase sub. d 1 | P51863 | 40302,8 | 6 | 23,60% |
|  | Nucleophosmin | Q61937 | 32560,3 | 4 | 22,90% |
|  | Sgta | Q8BJU0 | 34157,8 | 3 | 11,10% |
|  | Elongation factor 1-delta | P57776 | 31293,4 | 2 | 9,25% |
|  | Phospholipid scramblase 1 | Q9JJ00 | 35913,5 | 2 | 7,01% |
| | Adprhl2 | Q8CG72 | 39414,3 | 1 | 3,24% |
| 6A | eIF 4A-III | Q91VC3 | 46842,4 | 8 | 19,50% |
| 6B | eIF 4A-III | Q91VC3 | 46842,4 | 8 | 27,50% |
| | Adss2 | P46664 | 50140,8 | 4 | 11,00% |
| | Clp1 | Q99LI9 | 47629,1 | 2 | 4,71% |
| | Cathepsin D | P18242 | 44955 | 1 | 4,39% |
| | C-terminal-binding protein 1 | O88712 | 47744,7 | 1 | 2,49% |
| | Elongation factor 1-gamma | Q9D8N0 | 50061,3 | 1 | 2,97% |
| | EF-Tu mitochondrial | Q8BFR5 | 49399,2 | 1 | 2,65% |
| 7A | Protein NDRG1 | Q62433 | 43008,2 | 2 | 7,61% |
| 7B | Protein NDRG1 | Q62433 | 43008,2 | 3 | 11,90% |
| | TAR DNA-binding protein 43 | Q921F2 | 44547,5 | 3 | 9,90% |
| | Coronin-1A | O89053 | 50988,9 | 2 | 8,03% |
| | Creatine kinase B-type | Q04447 | 42714,1 | 2 | 7,35% |
| | Aldehyde dehydrogenase | P47738 | 56537,6 | 1 | 4,43% |
| | BRCA1-A complex subunit BRE | Q8K3W0 | 43560,1 | 1 | 2,61% |
| | Cathepsin D | P18242 | 44955 | 1 | 4,39% |
| | Eif3G | Q9Z1D1 | 35639 | 1 | 4,06% |
| | Plastin-2 | Q61233 | 70151,9 | 1 | 2,39% |
| 8A | Psma2 | P49722 | 25926,9 | 1 | 5,98% |
| 8B | Psma2 | P49722 | 25926,9 | 2 | 12,00% |
| | 40S ribosomal protein SA | P14206 | 32885,3 | 2 | 9,49% |
| | GAPDH | P16858 | 35828,1 | 1 | 4,20% |